\documentclass[12pt]{article}
\usepackage{graphicx}
\usepackage{amsmath}
\usepackage{amssymb}

\usepackage{amsmath}
\begin{document}

\title{Propagators of the Dirac fermions on spatially flat FLRW spacetimes}

\author{Ion I. Cot\u aescu \footnote{e-mail: i.cotaescu@e-uvt.ro}\\
{\small \it West University of Timi\c soara,}\\{\small \it V. P\^ arvan Ave. 4, RO-300223, Timi\c soara, Romania}}
%

\maketitle

\begin{abstract}
The general formalism of  the free Dirac fermions on spatially flat $(1+3)$-dimensional  Friedmann-Lema\^ itre-Robertson-Walker (FLRW)  spacetimes is developed in momentum representation. The mode expansions in terms of the fundamental spinors satisfying the charge conjugation and normalization conditions are used for deriving the structure of the anti-commutator matrix-functions and, implicitly, of the retarded, advanced, and Feynman fermion propagators. The principal  result is that the new type of integral representation we proposed recently in the de Sitter case can be applied to the Dirac fermions  in any spatially flat FLRW geometry.  Moreover,  the Dirac equation of the left-handed massless fermions can be analytically solved finding a general spinor solution and deriving the integral representations of the neutrino propagators.  It is shown that in the Minkowski flat spacetime our new integral representation is up to a change of variable just like the usual Fourier representation of the fermion propagators. The form of the Feynman propagator of the massive fermions on a spatially flat FLRW spacetime with a scale factor of Milne-type  is also outlined.

Pacs: 04.62.+v
\end{abstract}

\newpage

\section{Introduction}

In general relativity, the standard quantum field theory (QFT) based on perturbations and renormalization procedures was neglected paying more attention to alternative non-perturbative  methods as, for example, the cosmological creation of elementary particles of various spins \cite{P1,P2,S1}. The hope was to avoid thus the difficulties of the QFT arising especially on the FLRW manifolds  which evolve in time  \cite{BD}. In this context, the Dirac field  in the local charts with spherical coordinates of the FLRW manifolds was studied by many authors  which found that its time evolution is governed by a pair of time modulation functions that, in general, cannot be solved analytically since they satisfy equations of oscillators with variable frequencies  \cite{c1,c2,c3,c4,c5}.

In the simpler case of the spatially flat FLRW spacetimes, which are of actual interest in developing the  $\Lambda$CDM model, the symmetry under space translations allows one to consider plane waves instead of the spherical ones, obtaining quantum modes which have similar properties as in special relativity. Nevertheless, despite of this obvious advantage,  the  perturbative QFT  on spatially flat FLRW manifolds  is inchoate since  in the actual theory of propagators (or two-point functions) a suitable integral representation (rep.)  we need for calculating Feynman diagrams is missing. For this reason we would like to focus here on the general structure of the fermion propagators on spatially flat FLRW spacetime proposing a new type of integral rep. that may hold on these manifolds.

The propagators of the Dirac fermions are well-studied on the de Sitter expanding universe exploiting the fact that this is the expanding portion of the spacetime of maximal symmetry and negative (global) curvature \cite{SW} equipped with spatially flat FRLW or conformal local charts. These propagators were derived first by Candelas and Reine  integrating the Green equation  \cite{CR} and then by Koskma and Prokopec which solved the mode integrals  in a more general context of spacetimes of arbitrary dimensions approaching to the de Sitter one \cite{KP}. However, these propagators, depending explicitely on the Heaviside step functions, cannot be used in calculating Feynman diagrams without an integral rep. which should encapsulate the effect of these functions. This is why we proposed recently a new type of integral rep.  which takes over the effect of the step functions but is different from the Fourier integrals used  in special relativity \cite{CIR}. Thus we completed the theory of the free Dirac fermions on the de Sitter expanding universe  \cite{CD1,CD2,CD3} with  the integral rep. of the  Feynman propagators we need for calculating physical effects in our de Sitter quantum electrodynamics (QED) in Coulomb gauge \cite{CQED} using perturbations. 

Our principal objective in the present paper is to extend this formalism to the free Dirac fermions on any $(1+3)$-dimensional  spatially flat FLRW spacetime.  As mentioned, these manifolds have at least an isometry group including translations allowing us to chose the  momentum rep. where the fundamental solutions of the Dirac equation are plane waves whose time evolution is given by two modulation functions.   Under such circumstances, it is convenient to separate the orbital and spin parts of the fundamental spinors since in this manner we may impose the symmetry under charge conjugation and normalization without solving the pair of modulation functions.  After building the set of fundamental spinors which form an othonormal basis we derive the anti-commutator (acom.)  matrix-functions as mode integrals and define the retarded, advanced and Feynman propagators.  Furthermore, as an auxiliary result, we generalize the method of Ref. \cite{KP} showing that the mode integrals defining the acom. matrix-functions can be solved in terms of  only one integral involving the  modulation functions. In this framework, we may introduce naturally our new integral rep. of the fermion propagators in any spatially flat FLRW geometry, exploiting the method of the contour integrals \cite{BDR,Complex}.  We must stress that, in general,  this integral rep. is different from the usual Fourier rep. of the Feynman propagators in special relativity.  Moreover, since in the FLRW manifolds have conformally flat local charts where the fundamental spinors of the massless fermions  are just those of special relativity multiplied with a conformal factor,  we can write down the complete theory of the acom.  matrix-functions and propagators of the left-handed massless fermions (neutrinos). 

We obtain thus a coherent general formalism depending only on the pair of modulation functions that satisfy a simple system of differential equation, of oscillators with variable frequencies,  which can be solved in some concrete geometries. Here we restrict ourselves to  the classical examples of the Minkowski, de Sitter and a version of Milne-type spacetimes but there are many other cases in which this system can be analytically solved.  The first example is important since we have the opportunity to show that in the flat spacetime our integral rep. coincides, up to a change of integration variable, to the usual Fourier rep. of the fermion propagators in special relativity.  Finally, we revisit the de Sitter case which inspired this approach and deduce the form of the fundamental spinors on a spatially flat FLRW spacetime with a Milne-type scale factor outlining the form of the fermion propagators on this manifold.

We start in the second section presenting the general properties of the Dirac field on the spatially flat FLRW spacetime whose fundamental solutions comply with the orthonormalization condition and charge conjugation. We separate the orbital and spin parts of these spinors pointing out the modulation functions  which satisfy a simple system of differential equations whose prime integrals allow us to impose the normalization conditions. The acom.  matrix-functions and the related propagators are introduced in the third section where we obtain their general forms depending on a single integral involving the modulation functions. In the next section we consider the method of contour integrals for showing that our new integral rep. \cite{CIR} can be applied without restrictions. The propagators of the left-handed massless fermions are derived in the fifth section  by using the analytic solutions of the modulation functions. The next section is devoted to the classical examples:  Minkowski, de Sitter and Milne-type spacetimes. Therein we show that our integral rep. is equivalent to the Fourier rep. in Minkowski spacetime up to a simple change of the integration variable. After revisiting the de Sitter case we solve the modulation functions on the mentioned Milne-type spacetime giving the functions involved in the integral rep. of the Feynman propagator.  Finally, we present our concluding remarks.

\section{The free Dirac field}

The spatially flat FLRW manifolds have at least the isometry group $E(3)=T(3)\circledS SO(3)$, i. e.  the semidirect product between the group of the space translations, $T(3)$, and the  rotations group, $SO(3)$, where $T(3)$ is the invariant subgroup.  The $SO(3)$  symmetry can be preserved as a global one by using coordinates $x^{\mu}$ (labeled  by the natural indices $\mu,\nu,...=0,1,2,3 $) formed by the time, $t$, and Cartesian space coordinates, $x^i$  ($i,j,k...=1,2,3$),  for which we may use the vector notation ${\bf x}=(x^1,x^2,x^3)$. 
Here we consider two types of comoving charts \cite{BD}:  the standard FLRW charts  $\{t,{\bf x}\}$ with the proper time $t$, and  the conformal flat charts, $\{t_c,{\bf x}\}$, where we use the conformal time $t_c$. The FLRW geometry is given by a smooth scale factor $a(t)$ defining the conformal time as,
\begin{equation}
t_c=\int \frac{dt}{a(t)}~\to~  a(t_c)=a[t(t_c)]\,.
\end{equation}
and determining the line elements,  
\begin{eqnarray}
ds^2=g_{\mu\nu}(x)dx^{\mu}dx^{\nu}&=&dt^2-a(t)^2 d{\bf x}\cdot d{\bf x}\nonumber\\
&=&a(t_c)^2(dt_c^2-d{\bf x}\cdot d{\bf x})\,.
\end{eqnarray}

In these charts, the vector fields $e_{\hat\alpha}=e_{\hat\alpha}^{\mu}\partial_{\mu}$ defining the local orthogonal frames,  and the 1-forms $\omega^{\hat\alpha}=\hat e_{\mu}^{\hat\alpha}dx^{\mu}$ of the dual coframes are labeled by the local indices, $\hat\mu,\hat\nu,...=0,1,2,3$.  In a given tetrad gauge, the  metric tensor is expressed as $g_{\mu\nu}=\eta_{\hat\alpha\hat\beta}\hat e^{\hat\alpha}_{\mu}\hat e^{\hat\beta}_{\nu}$ where $\eta={\rm diag}(1,-1,-1,-1)$ is the Minkowski metric. In what follows we consider only the diagonal tetrad gauge defined as  
\begin{eqnarray}
&e_0=\partial_t=\frac{1}{a(t_c)}\,\partial_{t_c}\,,\qquad & \omega^0=dt=a(t_c)dt_c\,,\label{tetrad} \\
&~~~e_i=\frac{1}{a(t)}\,\partial_i=\frac{1}{a(t_c)}\,\partial_i\,, \qquad & \omega^i=a(t)dx^i=a(t_c)dx^i\,,
\end{eqnarray}
in order to preserve the global $SO(3)$ symmetry  allowing us to use systematically the  $SO(3)$ vectors. 

The $E(3)$ isometries of the spatially flat FLRW manifolds gives rise to six classical conserved quantities, the components of the momentum and angular momentum that can be seen as $SO(3)$ vectors. In quantum theory, the corresponding conserved operators,  which commute with the operators of the field equations \cite{ES,EPL}, are the momentum operator ${\bf P}$ of components $P^i=-i\partial_i$ and the angular momentum one ${\bf L}={\bf x}\land {\bf P}$ that  help us to construct the bases of the momentum or angular momentum reps..  Note that the conserved momentum is different from the covariant momentum satisfying the geodesic equation. 

In this tetrad-gauge, the massive Dirac field $\psi$ of mass $m$ and its Dirac adjoint $\bar{\psi}=\psi^+\gamma^0$ satisfy the field equations $(D_x-m) \psi (x)=0$ and, respectively, $\bar{\psi}(x)(\bar{D}_x-m)=0$ given by the Dirac operator 
\begin{equation}\label{ED}
D_x=i\gamma^0\partial_{t}+i\frac{1}{a(t)}\gamma^i\partial_i
+\frac{3i}{2}\frac{\dot{a}(t)}{a(t)}\gamma^{0}\,,
\end{equation}
and its adjoint 
\begin{equation}\label{ED1}
\bar{D}_x=-i\gamma^0\stackrel{\leftarrow}\partial_{t}-i\frac{1}{a(t)}\gamma^i\stackrel{\leftarrow}\partial_i
-\frac{3i}{2}\frac{\dot{a}(t)}{a(t)}\gamma^{0}\,,
\end{equation}
whose derivatives act to the left. These operators are expressed in terms of the Dirac $\gamma$-matrices  and function $a(t)$ and its derivative denoted as $\dot{a}(t)=\partial_ta(t)$. It is known that the terms of these operators depending on the Hubble function $\frac{\dot{a}}{a}$ can be removed at any time by substituting $\psi \to [a(t)]^{-\frac{3}{2}}\psi$. Similar results can be written in the conformal chart.

The general solution of the Dirac equation  may be written as a mode integral, 
\begin{eqnarray}
\psi(t,{\bf x}\,)& =& 
\psi^{(+)}(t,{\bf x}\,)+\psi^{(-)}(t,{\bf x}\,)\nonumber\\
& =& \int d^{3}p
\sum_{\sigma}[U_{{\bf p},\sigma}(x){\frak a}({\bf p},\sigma)
+V_{{\bf p},\sigma}(x){\frak b}^{\dagger}({\bf p},\sigma)]\,,\label{p3}
\end{eqnarray}
in terms of the fundamental spinors $U_{{\bf p},\sigma}$  and  $V_{{\bf p},\sigma}$ of positive and respectively negative frequencies which are plane waves solutions of the Dirac equation depending on the conserved momentum ${\bf p}$ and an arbitrary polarization $\sigma$. These spinors satisfy the eigenvalues problems $P^i U_{{\bf p},\sigma}(t,{\bf x})=p^i U_{{\bf p},\sigma}(t,{\bf x})$ and  $P^i V_{{\bf p},\sigma}(t,{\bf x})=-p^i V_{{\bf p},\sigma}(t,{\bf x})$ and form an orthonormal  basis being related  through the charge conjugation, 
\begin{equation}\label{chc}
V_{{\bf p},\sigma}(t,{\bf x})=U^c_{{\bf p},\sigma}(t,{\bf x}) =C\left[\bar{U}_{{\bf p},\sigma}(t,{\bf x})\right]^T \,, \quad C=i\gamma^2\gamma^0\,,
\end{equation}
(see the  Appendix A), and satisfying the orthogonality relations
\begin{eqnarray}
\langle U_{{\bf p},\sigma}, U_{{{\bf p}\,}',\sigma'}\rangle &=&
\langle V_{{\bf p},\sigma}, V_{{{\bf p}\,}',\sigma'}\rangle=
\delta_{\sigma\sigma^{\prime}}\delta^{3}({\bf p}-{\bf p}\,^{\prime})\label{ortU}\\
\langle U_{{\bf p},\sigma}, V_{{{\bf p}\,}',\sigma'}\rangle &=&
\langle V_{{\bf p},\sigma}, U_{{{\bf p}\,}',\sigma'}\rangle =0\,, \label{ortV}
\end{eqnarray}
with respect to the relativistic scalar product \cite{CD1}
\begin{equation}
\langle \psi, \psi'\rangle=\int d^{3}x
\sqrt{|g|}\,e^0_0\,\bar{\psi}(x)\gamma^{0}\psi(x)=\int d^{3}x\,
a(t)^{3}\bar{\psi}(x)\gamma^{0}\psi(x)\,, 
\end{equation}
where $g={\rm det}(g_{\mu\nu})$. Moreover,  this basis is supposed to be complete accomplishing the completeness condition  \cite{CD1}
\begin{eqnarray}
&&\int d^{3}p
\sum_{\sigma}\left[U_{{\bf p},\,\sigma}(t,{\bf x}\,)U^{+}_{{\bf p},\sigma}(t,{\bf x}\,^{\prime}\,)+V_{{\bf p},\sigma}(t,{\bf x}\,)V^{+}_{{\bf p},\sigma}(t,{\bf x}\,^{\prime}\,)\right] \nonumber\\
&&\hspace*{18mm}=a(t)^{-3}\delta^{3}({\bf x}-{\bf x}\,^{\prime})\,.\label{complet}
\end{eqnarray}
We obtain thus the orthonormal basis of the momentum rep.  in which   the particle $({\frak a},{\frak a}^{\dagger})$ and antiparticle (${\frak b},{\frak b}^{\dagger})$ operators  satisfy the canonical anti-commutation relations \cite{CD1,CdSquant}, 
\begin{equation}
\{{\frak a}({\bf p},\sigma),{\frak a}^{\dagger}({\bf p}\,\,^{\prime},\sigma^{\prime})\}=
\{{\frak b}({\bf p},\sigma),{\frak b}^{\dagger}({\bf p}\,\,^{\prime},\sigma^{\prime})\}=\delta_{\sigma\sigma^{\prime}}
\delta^{3}({\bf p}-{\bf p}\,^{\prime})\,,
\end{equation}
which guarantee that  the one-particle operators conserved via Noether theorem become just the generators of the corresponding isometries \cite{CdSquant}.

The general form of the fundamental spinors can be studied  in any spatially flat FLRW geometry,  exploiting the Dirac equation in momentum rep..   For our further  purposes, it is convenient to separate from the beginning the orbital part from the spin terms as  
\begin{eqnarray}
U_{\vec p,\sigma}(t,{\bf x})&=&[2\pi a(t)]^{-\frac{3}{2}}{e^{i{\bf p}\cdot{\bf x}}}{\cal U}_p(t)\gamma({\bf p})u_{\sigma}\label{U}\\
V_{\vec p,\sigma}(t,{\bf x})&=&[2\pi a(t)]^{-\frac{3}{2}}{e^{-i{\bf p}\cdot{\bf x}}}{\cal V}_p(t)\gamma({\bf p}) v_{\sigma}\label{V}
\end{eqnarray}
where we introduce the diagonal matrix-functions ${\cal U}_p(t)$ and ${\cal V}_p(t)$  which depend only on $t$ and $p=|{\bf p}|$, determining the time modulation of the fundamental spinors. 

The spin part is separated with the help of the nilpotent matrix 
\begin{equation}\label{gamp}
\gamma({\bf p})=\gamma^0-\frac{ {\gamma}^i {p^i}}{p}\,, 
\end{equation}
acting on the rest frame spinors of the momentum-spin basis that in the standard rep. of the Dirac matrices (with diagonal $\gamma^0$) read \cite{TH}
\begin{equation}\label{Rfspin}
u_{\sigma}=\left(
\begin{array}{c}
\xi_{\sigma}\\
0
\end{array}\right)\,,\quad
v_{\sigma}=-C \bar u_{\sigma}^T=\left(
\begin{array}{c}
0\\
\eta_{\sigma}
\end{array}\right)\,,
\end{equation}
since $\gamma({\bf p})^c=-\gamma({\bf p})$ in Eq. (\ref{chc}). When we desire to work in the mo\-mentum-helicity basis  we have to replace 
\begin{equation}
u_{\sigma}\to u_{\lambda}=\left(
\begin{array}{c}
\xi_{\lambda}({\bf p})\\
0
\end{array}\right)\,, \quad v_{\sigma}\to v_{\lambda}=
\left(
\begin{array}{c}
0\\
\eta_{\lambda}({\bf p})
\end{array}\right)
\end{equation} 
without changing the structure of the fundamental spinors (\ref{U}) and (\ref{V}).  The  Pauli spinors of the spin basis, $\xi_{\sigma}$ and $\eta_{\sigma}$, as well as those of the helicity basis,  $\xi_{\lambda}({\bf p})$ and  $\eta_{\lambda}({\bf p})$, are given in the Appendix A. They are normalized and satisfy the completeness relation (\ref{Pcom}) and a similar one for the helicity spinors. Consequently,  one can use the projector matrices \cite{BDR,TH}
\begin{eqnarray}
&&\pi_+=\sum_{\sigma}u_{\sigma}\bar{u}_{\sigma}=\sum_{\lambda}u_{\lambda}\bar{u}_{\lambda}=\frac{1+\gamma^0}{2}\,,\\
&&\pi_-=\sum_{\sigma}v_{\sigma}\bar{v}_{\sigma}=\sum_{\lambda}v_{\lambda}\bar{v}_{\lambda}=\frac{1-\gamma^0}{2}\,,
\end{eqnarray}
that form a complete system since $\pi_+\pi_-=0$ and $\pi_++\pi_-=1$. All these auxiliary quantities will be useful in the further calculations having simple calculation rules as, for example, $\gamma({\bf p})^2=0$, $~\gamma({\bf p})\gamma(-{\bf p})=2\gamma({\bf p})\gamma^0$, $\gamma({\bf p})\pi_{\pm}\gamma({\bf p})=\pm\gamma({\bf p})$, etc..    

The principal pieces are the diagonal matrix-functions determining the time modulation of the fundamental spinors which can be represented as
\begin{eqnarray}
{\cal U}_p(t)&=&\pi_+ u_p^+(t)+\pi_-u_p^-(t)\,,\\
{\cal V}_p(t)&=&\pi_+ v_p^+(t)+\pi_-v_p^-(t)\,,
\end{eqnarray}
in terms of  the time modulation functions $u_p^{\pm}(t)$ and $v_p^{\pm}(t)$.  These  matrix-functions have the obvious properties as $\bar{\cal U}_p={\cal U}_p^+={\cal U}_p^*$ and similarly for ${\cal V}_p$. Moreover, when $U_{\vec p,\sigma}(t,{\bf x})$ and $V_{\vec p,\sigma}(t,{\bf x})$ satisfy the Dirac equation then we find the remarkable identities,
\begin{eqnarray}
&&a(t)(D_x+m)\left[{e^{i{\bf p}\cdot{\bf x}}}\,a(t)^{-\frac{3}{2}}\gamma^5{\cal U}_p(t)\gamma^5\right]={e^{i{\bf p}\cdot{\bf x}}}\,a(t)^{-\frac{3}{2}}{\cal U}_p(t) p \gamma({\bf p})\,,\label{id1}\\
&&a(t)(D_x+m)\left[{e^{-i{\bf p}\cdot{\bf x}}}\,a(t)^{-\frac{3}{2}}\gamma^5{\cal V}_p(t)\gamma^5\right]=-{e^{-i{\bf p}\cdot{\bf x}}}\,a(t)^{-\frac{3}{2}}{\cal V}_p(t) p \gamma({\bf p})\,,\nonumber\\ \label{id2}
\end{eqnarray}
that can be used in further applications.  

The next step is to derive the differential equations of the modulation  functions $u_p^{\pm}$ and $v^{\pm}_p$ in the general case of $m \not= 0$  by substituting Eqs. (\ref{U}) and (\ref{V}) in the Dirac equation.  Then, after a few manipulation, we find the systems of the first order differential equations
\begin{eqnarray}
a(t)\left(i\partial_t\mp m\right)u_p^{\pm}(t)&=&{p}\,u_p^{\mp}(t)\,,\label{sy1}\\
a(t)\left(i\partial_t \mp m\right)v_p^{\pm}(t)&=&-{p}\,v_p^{\mp}(t)\,,\label{sy2}
\end{eqnarray}
in the chart with the proper time or the equivalent system in the conformal chart,
\begin{eqnarray}
\left[i\partial_{t_c}\mp m\, a(t_c)\right]u_p^{\pm}(t_c)&=&{p}\,u_p^{\mp}(t_c)\,,\label{sy1c}\\
\left[i\partial_{t_c} \mp m\, a(t_c)\right]v_p^{\pm}(t_c)&=&-{p}\,v_p^{\mp}(t_c)\,,\label{sy2c}
\end{eqnarray}
which govern the time modulation of the free Dirac field on any spatially flat FLRW manifold. Note that these equations are similar to those of the modulation functions of the spherical modes \cite{c1,c2,c3,c4,c5} but depending on different integration constants.

The solutions of these systems depend on other integration constants that must be selected according to the charge conjugation (\ref{chc}) which requires to have   
${\cal V}_p={\cal U}^c_p=C{\cal U}_p^*C^{-1}=\gamma^5 {\cal U}_p^*\gamma^5$ 
leading to the mandatory condition
\begin{equation}\label{VU}
v_p^{\pm}(t)=\left[u_p^{\mp}(t)\right]^*\,.
\end{equation}
The remaining normalization constants can be fixed since the prime integrals of the systems (\ref{sy1}) and (\ref{sy2}), 
$\partial_t (|u_p^+|^2+|u_p^-|^2)=\partial_t (|v_p^+|^2+|v_p^-|^2)=0$, 
allow us to impose the normalization conditions
\begin{equation}
|u_p^+|^2+|u_p^-|^2=|v_p^+|^2+|v_p^-|^2 =1 \label{uuvv}\\
\end{equation}
which guarantee that Eqs.  (\ref{ortU}) and (\ref{ortV}) are accomplished. Hereby we find  the calculation rules 
${\rm Tr}({\cal U}_p{\cal U}_p^*)={\rm Tr}({\cal V}_p{\cal V}_p^*)=2$ 
resulted from Eqs.  (\ref{uuvv}) and ${\rm Tr}(\pi_{\pm})=2$.

A special problem is that of the rest frame where ${\bf p}=0$ since here the matrix (\ref{gamp}) is not defined. Nevertheless, the Dirac equation in momentum rep.  can be solved analytically in this case giving the fundamental spinors of the rest frame,
\begin{eqnarray}
U_{0,\sigma}(t,{\bf x})&=&[2\pi a(t)]^{-\frac{3}{2}}e^{-i mt}u_{\sigma}\,,\label{Ur}\\
V_{0,\sigma}(t,{\bf x})&=&-[2\pi a(t)]^{-\frac{3}{2}}e^{i m t} v_{\sigma}\,,\label{Vr}
\end{eqnarray} 
which depend on the rest energy $E_0=m$ and the rest frame spinors $u_{\sigma}$ and $v_{\sigma}$. 

We have determined here the structure of the fundamental spinors on spatially flat FLRW spacetimes up to a pair of modulation  functions, $u_p^{\pm}(t)$ or $u_p^{\pm}(t_c)$, which  can be found by integrating  the systems (\ref{sy1}) or (\ref{sy1c})  in each particular case separately and imposing the normalization condition (\ref{uuvv}). Thus we obtain the normalized spinors of positive frequencies while the negative frequency ones have to be derived by using the identities (\ref{VU}).

\section{Propagators}

In the quantum theory of fields it is important to study  the Green functions related to the total or partial acom. matrix-functions of positive or negative frequencies \cite{CD1},
\begin{equation}
{S}^{(\pm)}(t,t^{\prime},{\bf x}-{\bf x}\,^{\prime}\,)=-i\{\psi^{(\pm)}(t,{\bf x})\,,\bar{\psi}
^{(\pm)}(t^{\prime},{\bf x}\,^{\prime}\,)\}\,,\label{Spm}
\end{equation}
which satisfy the Dirac equation in both sets of variables,
\begin{equation}
(D_x - m){S}^{(\pm)}(t,t^{\prime},{\bf x}-{\bf x}\,^{\prime}\,)={S}^{(\pm)}(t,t^{\prime},{\bf x}-{\bf x}\,^{\prime}\,)(\bar{D}_{x'}-m)=0\,.
\end{equation}
The total acom.  matrix-function \cite{CD1}
\begin{eqnarray}
{S}(t,t^{\prime},{\bf x}-{\bf x}\,^{\prime}\,)&=&-i\{\psi(t,{\bf x}\,),\bar{\psi}(t',{\bf x}\,^{\prime}\,)\}\nonumber\\
&=&{S}^{(+)}(t,t^{\prime},{\bf x}-{\bf x}\,^{\prime}\,)+{S}^{(-)}(t,t^{\prime},{\bf x}-{\bf x}\,^{\prime}\,)\,,\label{Stot}
\end{eqnarray} 
has similar properties and, in addition, satisfy the equal-time condition
\begin{equation}
{S}(t,t,{\bf x}-{\bf x}\,^{\prime}\,)=-i\gamma^0 a(t)^{-3}\delta^{3}({\bf x}-{\bf x}\,^{\prime})
\end{equation}
resulted from Eq. (\ref{complet}). 

These matrix-functions allow one to introduce the Green functions corresponding to asymptotic initial conditions without solving the Green equation. These are the retarded (R) and advanced (A) Green functions,
\begin{eqnarray}
S_R(t,t',{\bf x}-{\bf x}\,^{\prime}\,)&=&\theta(t-t')S(t,t',{\bf x}-{\bf x}\,^{\prime}\,)\label{SR}\\
S_A(t,t',{\bf x}-{\bf x}\,^{\prime}\,)&=&-\theta(t'-t)S(t,t',{\bf x}-{\bf x}\,^{\prime}\,)\label{SA}
\end{eqnarray}
which have the property
\begin{equation}
S_R(t,t',{\bf x}-{\bf x}\,^{\prime}\,)-S_A(t,t',{\bf x}-{\bf x}\,^{\prime}\,)={S}(t,t^{\prime},{\bf x}-{\bf x}\,^{\prime}\,)\,,
\end{equation}
that may be used in the reduction formalism. The Feynman propagator,
\begin{eqnarray}
&&S_{F}(t,t^{\prime},{\bf x}-{\bf x}\,^{\prime})=
-i\langle0|T[\psi(x)\bar{\psi}(x^{\prime})]|0\rangle\nonumber\\
 &&= \theta(t-t^{\prime})S^{(+)}(t,t^{\prime},{\bf x} -{\bf x}\,^{\prime}\,)-\theta(t^{\prime}-t)S^{(-)}(t,t^{\prime},{\bf x}-{\bf x}\,^{\prime})\,,\label{SF}
\end{eqnarray}
is the Green function with a causal structure describing the propagation of the particles and antiparticles \cite{BDR}.  Obviously, all these Green functions satisfy the Green equation that in the FLRW chart has the form \cite{CD1},
\begin{eqnarray}
(D_x-m)S_{\substack{F\\R\\A}}(t,t^{\prime},{\bf x}-{\bf x}\,^{\prime})&=&S_{\substack{F\\R\\A}}(t,t^{\prime},{\bf x}-{\bf x}\,^{\prime})(\bar{D}_{x'}-m)\nonumber\\
&=&a(t)^{-3} \delta^{4}(x-x\,^{\prime})\,. \label{p8}
\end{eqnarray}
We remind the reader that this equation has an infinite set of solutions corresponding to various initial conditions but here we focus only on the Green functions $S_R$, $S_A$ and $S_F$ which will be called propagators in what follows.  

The propagators are related to the acom.  matrix-functions which  can be written  in terms of the fundamental spinors   (\ref{U}) and (\ref{V}) as two similar mode integrals,
\begin{eqnarray}
&&i{S}^{(+)}(t,t^{\prime},{\bf x}-{\bf x}\,^{\prime}\,)=\sum_{\sigma}\int d^3p\, U_{{\bf p},\sigma}(t,{\bf x}\,)\bar{U}_{{\bf p},\sigma}(t\,^{\prime},{\bf x}\,^{\prime}\,) \nonumber\\
&&~~~=n(t,t')\int d^3p\, e^{i{\bf p}\cdot({\bf x}-{\bf x}')}{\cal U}_p(t)\gamma({\bf p})[{\cal U}_p(t')]^*\,,\label{Splus}\\
&&i{S}^{(-)}(t,t^{\prime},{\bf x}-{\bf x}\,^{\prime}\,)=\sum_{\sigma}\int d^3p\, V_{{\bf p},\sigma}(t,{\bf x}\,)\bar{V}_{{\bf p},\sigma}(t\,^{\prime},{\bf x}\,^{\prime}\,) \nonumber\\
&&~~~= n(t,t')\int d^3p\, e^{i{\bf p}\cdot({\bf x}-{\bf x}')}{\cal V}_p(t)\gamma(-{\bf p})[{\cal V}_p(t')]^*\,, \label{Smin}
\end{eqnarray} 
after denoting
\begin{equation}
n(t,t')={[4\pi^2 a(t) a(t')]^{-\frac{3}{2}}}\,,
\end{equation}
and changing ${\bf p}\to-{\bf p}$ in the last integral (without affecting ${\cal V}_p$ which depend only on $p$). 

This is the starting point for finding how these matrix-functions and implicitly the propagators depend on coordinates, after solving the mode integrals.  In order to do this, we follow the method of Ref. \cite{KP} introducing  the new matrix-functions $\Sigma^{(\pm)}$ defined as, 
\begin{equation}\label{SSig}
{S}^{(\pm)}(t,t^{\prime},{\bf x}-{\bf x}\,^{\prime}\,)=a(t)(D_x+m){\Sigma}^{(\pm)}(t,t^{\prime},{\bf x}-{\bf x}\,^{\prime}\,)\,.
\end{equation}
These can be related to the adjoint matrix-functions,
\begin{equation}
{\bar\Sigma}^{(\pm)}(t,t^{\prime},{\bf x}-{\bf x}\,^{\prime}\,)=\gamma^5 {\Sigma}^{(\pm)}(t,t^{\prime},{\bf x}-{\bf x}\,^{\prime}\,)\gamma^5\,.
\end{equation} 
which satisfy the adjoint relations  
\begin{equation}\label{SSig1}
{S}^{(\pm)}(t,t^{\prime},{\bf x}-{\bf x}\,^{\prime}\,)={\bar\Sigma}^{(\pm)}(t,t^{\prime},{\bf x}-{\bf x}\,^{\prime}\,)(\bar{D}_x+m)a(t')\,,
\end{equation}
given by the adjoint operator (\ref{ED1}). Furthermore,  we have to show that these new matrix-functions  have simpler forms depending in fact only on an integral involving the modulation functions that reads
\begin{eqnarray}
I(t,t',{\bf x})&=&\int \frac{d^3p}{p}\, e^{i{\bf p}\cdot{\bf x}}u_p^-(t)[u_p^+(t')]^*\nonumber\\
&=&
\frac{4\pi}{|{\bf x}|}\int_{0}^{\infty} dp\, u_p^-(t)[u_p^+(t')]^*\sin p|{\bf x}|\,,\label{Itt}
\end{eqnarray}
where we use the argument ${\bf x}$ instead of ${\bf x}-{\bf x}'$.  Indeed, according to Eqs. (\ref{id1}), (\ref{id2}) and (\ref{SSig}), after a little calculation, we obtain the definitive expressions
\begin{eqnarray}
i{\Sigma}^{(+)}(t,t^{\prime},{\bf x}) &=& n(t,t')\int \frac{d^3p}{p}\, e^{i{\bf p}\cdot{\bf x}}\gamma^5{\cal U}_p(t)\gamma^5[{\cal U}_p(t')]^*\nonumber\\
&=&n(t,t')\int \frac{d^3p}{p}\,  e^{i{\bf p}\cdot{\bf x}}
\left\{\pi_+ u_p^-(t)[u_p^+(t')]^*+\pi_- u_p^+(t)[u_p^-(t')]^*\right\}\nonumber\\
&=&n(t,t')\left[ \pi_+I(t,t',{\bf x})+\pi_-I(t',t,{\bf x})^*\right]\,, \label{Splus1}\\
i{\Sigma}^{(-)}(t,t^{\prime},{\bf x})& =& - n(t,t')\int \frac{d^3p}{p}\, e^{i{\bf p}\cdot{\bf x}}\gamma^5{\cal V}_p(t)\gamma^5[{\cal V}_p(t')]^*\nonumber\\
&=&-n(t,t')\int \frac{d^3p}{p}\, e^{i{\bf p}\cdot{\bf x}}
\left\{\pi_+ u_p^-(t')[u_p^+(t)]^*+\pi_- u_p^+(t')[u_p^-(t)]^*\right\}\nonumber\\
&=&-n(t,t')\left[ \pi_+I(t',t,{\bf x})+\pi_-I(t,t',{\bf x})^*\right]\,,\label{Smin1}
\end{eqnarray}
which depend only on the integral (\ref{Itt}) which will be referred here as the basic integral.  In the case of the massive Dirac field this integral has to be calculated in each particular case separately after deriving the modulation functions $u_p^{\pm}$. Similar expressions can be obtained in the conformal chart by changing $t \to t_c$ and $a(t)\to a(t_c)$.
 
Hereby we observe that the matrix-functions $\Sigma^{(\pm)}$ are related each other as
\begin{equation}\label{SiSi}
{\Sigma}^{(-)}(t,t^{\prime},{\bf x})=-{\Sigma}^{(+)}(t^{\prime},t,{\bf x})\,,
\end{equation} 
and are parity-invariant,  
\begin{equation}
{\Sigma}^{(\pm)}(t,t^{\prime},-{\bf x})={\Sigma}^{(\pm)}(t,t^{\prime},{\bf x})\,, 
\end{equation}
since the basic integral (\ref{Itt}) depends only on $|{\bf x}|$.

\section{Integral representations}

The propagators  (\ref{SR}), (\ref{SA}) and (\ref{SF}) depending explicitely on the Heaviside step functions cannot be used in concrete calculations involving time integrals as, for example, those of the Feynman diagrams.  In Minkowski spacetime this problem is solved by representing these propagators as four dimensional Fourier integrals which  take over the effects of the Heaviside functions according to the well-know method of the contour integrals \cite{BDR}. In this manner one obtains a suitable integral rep. of the Feynman propagators that can be used for performing calculations in  momentum rep..

In the spatially flat FLRW spacetimes we also have a momentum rep. but we cannot apply the method of Fourier transforms since in these geometries the propagators are functions of two separated variables, $t$ and $t'$, instead of the unique variable $t-t'$ of the flat case. This means that we might consider a double Fourier transform which is unacceptable from the point of wiev of the actual quantum theory. Therefore,  we must  look for an alternative integral rep. based on  the method of the contour integrals  \cite{BDR} but avoiding the mentioned Fourier transform. 

For this purpose we need to introduce a new variable of integration. This can be done in a natural manner observing that the systems (\ref{sy1c}) and (\ref{sy2c}) may be seen as an unique system, 
\begin{equation}\label{sys}
\left[i\partial_{t_c}\mp m\, a(t_c)\right]w_s^{\pm}(t_c)={s}\,w_s^{\mp}(t_c)\,,
\end{equation}
depending on the continuous parameter $s$ which may play the role of the desired new variable of integration. The functions $w_s^{\pm}$  whose restrictions are just
\begin{equation}
w_{s=p}^{\pm}(t_c)=u_p^{\pm}(t_c)\,, \quad w_{s=-p}^{\pm}(t_c)=v_p^{\pm}(t_c)\,,
\end{equation}
have similar properties as  $u_p^{\pm}$ and $v_p^{\pm}$, being symmetric with respect to the charge conjugation,  
\begin{equation}
w_{-s}^{\pm}(t_c)=\left[w_s^{\mp}(t_c)\right]^*\,,
\end{equation}
and satisfying the normalization condition
\begin{equation}
\left| w_s^+(t_c)\right|^2+\left| w_{s}^-(t_c)\right|^2=1\,, \qquad \forall s\in{\Bbb R}\,,
\end{equation}
allowed by the prime integral $\partial_{t_c}\{\left| w_s^+(t_c)\right|^2+\left| w_{s}^-(t_c)\right|^2\}=0$ of the system (\ref{sys}).
With these functions  we construct the diagonal matrix-function 
\begin{equation}
{\cal W}_s(t_c)=\pi^+w^+_s(t_c)+\pi^-w^-_s(t_c)\,,
\end{equation}
giving ${\cal U}_p={\cal W}_{s=p}$ and ${\cal V}_p={\cal W}_{s=-p}$ and having  the obvious properties 
\begin{equation}
{\cal W}_{-s}=\gamma^5 {\cal W}_s^*\gamma^5\,,\quad {\rm Tr}({\cal W}_s{\cal W}_s^*)=2\,,
\end{equation}
similar to  (\ref{VU}) and (\ref{uuvv}). 

{ \begin{figure}
  \centering
    \includegraphics[scale=0.40]{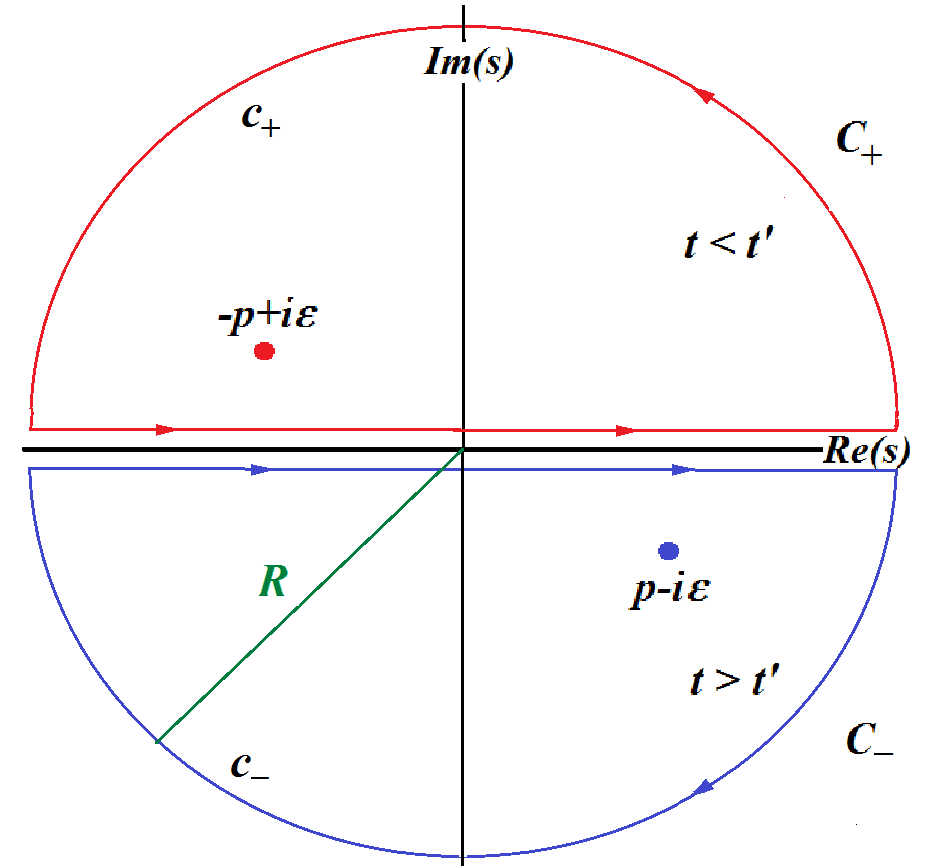}
    \caption{The contours of integration in the complex $s$-plane, $C_{\pm}$, are the limits of the pictured ones for $ R\to \infty$.}
  \end{figure}}

With these preparations we may propose the general integral rep.
\begin{eqnarray}
{S}_F(t_c,t_c^{\prime},{\bf x})
&=& \frac{1}{8\pi^4[a(t_c) a(t_c')]^{\frac{3}{2}}}\nonumber\\
&\times&\int {d^3p}\,{e^{i{\bf p}\cdot{\bf x}}}\int_{-\infty}^{\infty}ds\,  {\cal W}_{s}(t_c)\frac{\gamma^0s-\gamma^i p^i}{s^2-p^2+i\epsilon}\left[{\cal W}_{s}(t_c')\right]^*\,,\label{SF1}
\end{eqnarray} 
which encapsulates the effect of the Heaviside step functions that enter in the structure of the Feynman propagator (\ref{SF}). This formula can be written at any time in terms of proper times, by changing $t_c\to t$ and $a(t_c)\to a(t)$, since this integral rep. is independent on the chart we chose as long as we integrate with respect to the conserved momentum ${\bf p}$ and the associated new variable $s$.

The main task is to prove that this integral rep. gives just the Feynman propagator (\ref{SF}). In order to do this we must solve  the last integral of Eq. (\ref{SF1}) denoted now as  
\begin{equation}
{\cal I}(t_c,t_c')=\int_{-\infty}^{\infty}ds\,M(s,t_c,t_c')\,.
\end{equation}
For  large values of $|s|$ the system (\ref{sys}) may be approximated neglecting the mass terms such that we obtain the asymptotic solutions
\begin{equation}
w_s^{\pm}(t_c)\sim\frac{1}{\sqrt{2}}e^{-i s t_c}
\end{equation} 
which determines the behavior  
\begin{equation}
M(s,t_c,t_c')\sim \frac{\gamma^0s-\gamma^i p^i}{s^2-p^2+i\epsilon}\,e^{-is(t_c-t_c')}\,, 
\end{equation}
allowing us to estimate the integrals on the semicircular parts, $c_{\pm}$, of the contours pictured in Fig. 1, according to Eq. (3.338-6) of Ref. \cite{NIST}, as
\begin{equation}
\int_{c_{\pm}}ds\,M(s,t_c,t_c')\sim I_0[\pm R(t_c-t_c')]\sim \frac{1}{\sqrt{R}}\,e^{\pm R(t_c-t_c')}\,,
\end{equation}
since the modified Bessel function $I_0$ behaves as in  the first of Eqs. (\ref{Km0}). In the limit of $R\to \infty$ the contribution of the semicircle $c_+$ vanishes for $t_c'>t_c$ while those of the semicircle $c_-$ vanishes for $t_c>t_c'$. Therefore, the integration along the real $s$-axis  is equivalent with the following contour integrals
\begin{equation}
{\cal I}(t_c,t_c')=\left\{
\begin{array}{lll}
\int_{\small C_+}ds\,M(s,t_c,t_c')={\cal I}_+(t_c,t_c')&{\rm for}& t_c<t_c'\\
\int_{\small C_-}ds\,M(s,t_c,t_c')={\cal I}_-(t_c,t_c')&{\rm for}&t_c>t_c'
\end{array}\right. \,,\nonumber
\end{equation} 
where the contours $C_{\pm}$ are the limits for $R\to \infty$ of those of Fig. 1. Then we  may apply  the Cauchy's theorem \cite{Complex}, 
\begin{equation}
{\cal I}_{\pm}(t_c,t_c')=\pm 2\pi i \left.{\rm Res}\left[M(s,t_c,t_c')\right]\right|_{s=\mp p\pm i\epsilon}\,,
\end{equation}
taking into account that in the simple poles at $s=\pm p\mp i\epsilon$ we have the residues
\begin{equation}
\left.{\rm Res}\left[M(s,t,t')\right]\right|_{s=\pm p\mp i\epsilon}=\frac{p}{2}{\cal W}_{\pm p}(t_c)\gamma(\pm{\bf p})[{\cal W}_{\pm p}(t'_c)]^*\,.
\end{equation}
Consequently,  the integral ${\cal I}_-(t_c,t_c')$ gives the first term of the Feynman propagator (\ref{SF}) while the integral ${\cal I}_+(t_c,t_c')$ yields its second term, proving that the integral rep. (\ref{SF1}) is correct.  

The other propagators, $S_A$ and $S_R$, can be represented in a similar manner by changing the positions of the poles as in the flat case \cite{BDR},
\begin{eqnarray}
{S}_{\substack{R\\
A}}(t_c,t_c^{\prime},{\bf x})
&=& \frac{1}{8\pi^4[a(t_c) a(t_c')]^{\frac{3}{2}}}\nonumber\\
&\times&\int {d^3p}\,{e^{i{\bf p}\cdot{\bf x}}}\int_{-\infty}^{\infty}ds\,  {\cal W}_{s}(t_c)\frac{\gamma^0s-\gamma^i p^i}{(s\pm i\epsilon)^2-p^2}\left[{\cal W}_{s}(t_c')\right]^*\,,\label{SRA}
\end{eqnarray}  
but in our integral rep. instead of the Fourier one.

\section{Neutrino propagators}

An important particular case is of the left-handed massless neutrinos for which we must consider the limit to $m\to 0$ of the left-handed projection obtained with the help of the projector $P_L$ defined in Eqs. (\ref{PLR}). Fortunately, when $m=0$ the Dirac equation becomes conformaly covariant and, consequently, this can be solved analytically in any FLRW spacetime. More specific, we see that the systems  (\ref{sy1c}) and (\ref{sy2c}) in the conformal chart become independent on $a(t_c)$ giving the simple normalized solutions 
\begin{eqnarray}
&&u_+(t_c)=u_-(t_c)=\frac{1}{\sqrt{2}}\,e^{-ipt_c}\,, \label{u0v0}\\
 &&v_+(t_c)=v_-(t_c)=\frac{1}{\sqrt{2}}\,e^{ipt_c}\,,\label{u0v01}
\end{eqnarray}
which satisfy the condition (\ref{VU}).

In other respects, in the massless case the rest frame cannot be defined such that we must give up the momentum-spin basis considering the momentum-helicity basis in which the spin is projected along the momentum direction (as in the Appendix A). In addition. if we intend to separate the left-handed part it is convenient to consider the chiral rep. of the Dirac matrices (with diagonal $\gamma^5$)  in which the chiral projectors  $P_L$ and $P_R$  have diagonal forms.  Then the fundamental spinors of the left-handed massless Dirac field (neutrino) can be written as  \cite{CD1},
\begin{eqnarray}
U^0_{{\bf p},\lambda}(t_c,{\bf x})&=&
\lim_{m\to 0} \frac{1-\gamma^5}{2}\, U_{{\bf p},\lambda}(t_c,{\bf x})
\nonumber\\
&=&\left[{2\pi}\,a(t_c)\right]^{-3/2}
\left(
\begin{array}{c}
(\frac{1}{2}-\lambda) \xi_{\lambda}({\bf p})\\
0
\end{array}\right)
\,e^{-ipt_c+i{\bf p}\cdot{\bf x}}\,, \label{n1}\\
V^0_{{\bf p},\lambda}(t_c,{\bf x})&=&
\lim_{m\to 0} \frac{1-\gamma^5}{2}\,V_{{\bf p},\lambda}(t_c,{\bf x})
\nonumber\\
&=&\left[{2\pi}\,a(t_c)\right]^{-3/2}
\left(
\begin{array}{c}
(\frac{1}{2}+\lambda)\eta_{\lambda}({\bf p})\\
0
\end{array}\right)
\,e^{ipt_c-i{\bf p}\cdot{\bf x}}\,.  \label{n2}
\end{eqnarray}
These solutions are just those of the Minkowski spacetime  multiplied with the conformal factor $\left[{2\pi}\,a(t_c)\right]^{-3/2}$. Therefore, these have the only non-vanishing components  either of positive frequency and $\lambda=-\frac{1}{2}$ or of negative frequency and $\lambda=\frac{1}{2}$, as in the Minkowski spacetime.   

Now we can calculate the basic integral (\ref{Itt}) by using the solutions (\ref{u0v0}) and introducing the small $\epsilon>0$ for assuring its convergence. Thus  we obtain   
\begin{eqnarray}
I_{\epsilon}^0(t_c,t'_c,{\bf x})&=&\frac{2\pi}{|{\bf x}|}\int_0^{\infty} dp\, e^{-ip(t_c-t'_c-i\epsilon)}\sin p|{\bf x}|\nonumber\\
&=&\frac{2\pi}{|{\bf x}|^2-(t_c-t'_c-i\epsilon)^2}\,,
\end{eqnarray}
observing that 
\begin{equation}
I_{\epsilon}^0(t_c,t'_c,{\bf x})^*=I_{\epsilon}^0(t'_c,t_c,{\bf x})=I_{-\epsilon}^0(t_c,t'_c,{\bf x})\,.
\end{equation}
Therefore, we may write the closed expression,
\begin{equation}
i{\Sigma}^{(\pm)}_{0\,\epsilon}(t_c,t_c^{\prime},{\bf x})= \pm (2\pi)^{-2}\left[a(t_c) a(t'_c)\right]^{-\frac{3}{2}}
\left[|{\bf x}|^2-(t_c-t'_c\mp i\epsilon)^2\right]^{-1}\,,
\end{equation}
On the other hand, by using the solutions (\ref{n1}) and (\ref{n2}) and the projectors defined in the Appendix A, after a few manipulation, we may put  the mode integrals of the matrix-functions (\ref{Spm}) in the form
\begin{eqnarray}
i{S}^{(\pm)}_0(t_c,t_c^{\prime},{\bf x})
&=& n(t_c,t_c') \int d^3p\, \left(
\begin{array}{cc}
0&P_{\mp \frac{1}{2}}\\
0&0
\end{array}\right)
e^{\pm i{\bf p}\cdot{\bf x})\mp i p(t_c-t_c')}\nonumber\\
&=&n(t_c,t_c')\frac{1-\gamma^5}{2}\int d^3p\, \frac{\gamma(\pm{\bf p})}{2}e^{\pm i{\bf p}\cdot{\bf x}\mp i p(t_c-t_c')}\,,\label{S0pm}
\end{eqnarray}
since here we work with the chiral rep. of the Dirac matrices (with diagonal $\gamma^5$). Hereby we obtain the chiral projection giving the definitive form of the acom.  matrix-functions of the left-handed neutrinos,
\begin{equation}\label{SSig1}
i{S}^{(\pm)}_{0\,\epsilon}(t_c,t_c^{\prime},{\bf x})=\frac{1-\gamma^5}{2}\left[a(t_c)D_{x_c} {\Sigma}^{(\pm)}_{0\,\epsilon}(t_c,t'_c,{\bf x}\,^{\prime}\,)\right]\,,
\end{equation} 
in any FLRW geometry. This result can be rewritten at any time in the FLRW chart by changing the time variable $t_c\to t$ and $a(t_c)\to a(t)$.

Now the Feynman propagator of  the left-handed neutrinos  can be derived as  
\begin{equation}
{S}^0_{F}(t_c,t_c^{\prime},{\bf x})=\lim_{m\to 0}\frac{1-\gamma^5}{2}\, {S}_{F}(t_c,t_c^{\prime},{\bf x})\,,
\end{equation}
taking into account that for $m=0$ we may use the particular functions $K_{\frac{1}{2}}$ given by the last of Eqs. (\ref{Km0}). Thus we arrive at the final result
\begin{eqnarray}
{S}^0_{F}(t_c,t_c^{\prime},{\bf x})
&=& \frac{1}{(2\pi)^4[a(t_c) a(t_c')]^{\frac{3}{2}}}\nonumber\\
&\times&\int {d^3p}\int_{-\infty}^{\infty}ds\,\frac{1-\gamma^5}{2}\frac{\gamma^0s-\gamma^i p^i}{s^2-p^2+i\epsilon}\,e^{i{\bf p}\cdot{\bf x}-is(t_c-t_c')}\,.\label{SF2} 
\end{eqnarray}
which is just the Fourier integral of the Feynman propagator of the left-handed neutrinos in  Minkowski spacetime multiplied with the conformal factor $[a(t_c)c(t'_c)]^{-\frac{3}{2}}$.

\section{Classical examples}

The above presented approach can be applied easily to any particular case. For doing so we have to integrate one of the the systems (\ref{sy1}) or  (\ref{sy1c}) for finding the modulation functions $u_p^{\pm}$ giving the spinors of positive frequencies (\ref{U}).  Then by using Eq. (\ref{VU}) we obtain the functions $v_p^{\pm}$ we need for building the spinors of negative frequency  (\ref{V}). The next step is to calculate the basic integral (\ref{Itt}) for finding the closed forms of the acom. matrix functions. Finally, we have to write down the integral reps. of the propagators (\ref{SF1}) and (\ref{SRA}) with the help of the functions $w_s^{\pm}$ deduced from the system (\ref{sys}).

Let us see how this method works in the case of the classical examples, the Minkowski, de Sitter and a new Milne-type spacetimes.

\subsection{Minkowski spacetime}

The Minkowski spacetime is the simplest example with  $t_c=t$ and $a(t)=a(t_c)=1$ such that the solutions of the systems (\ref{sy1}) and (\ref{sy2}) which satisfy the conditions (\ref{VU}) and  (\ref{uuvv}) read
\begin{eqnarray}
u_p^{\pm}(t)&=&\sqrt{\frac{E(p)\pm m}{2 E(p)}}\,e^{-i E(p) t}\label{UMin}\\
v_p^{\pm}(t)&=&\sqrt{\frac{E(p)\mp m}{2 E(p)}}\,e^{i E(p) t}\label{VMin}
\end{eqnarray}
where $E(p)=\sqrt{p^2 +m^2}$. Thus we recover the standard fundamental spinors of the Dirac theory on Minkowski spacetime \cite{BDR}.

Another familiar result we recover is the form of the matrix-functions $\Sigma^{(\pm)}$ which now can be derived directly by substituting the functions (\ref{UMin}) in the basic integral (\ref{Itt}) where, in addition, we introduce a small $\epsilon>0$ for assuring the convergence. Then, according to Eq. (3.961-1) of Ref. \cite{GR}, we obtain the closed form
\begin{eqnarray}
&& I_{\epsilon}(t,t',{\bf x})=\frac{2\pi}{|{\bf x}|}\int dp \frac{p}{E(p)}e^{-i E(p)(t-t'-i\epsilon)}\sin p |{\bf x}|\nonumber\\
&&~~~~~~=\frac{2\pi m}{\sqrt{|{\bf x}|^2-(t-t'-i\epsilon)^2}} K_2\left(m\sqrt{|{\bf x}|^2-(t-t'-i\epsilon)^2}\right)
\end{eqnarray}
which satisfy 
\begin{equation}
I_{\epsilon}(t,t',{\bf x})=I_{-\epsilon}(t',t,{\bf x})=I_{\epsilon}(t',t,{\bf x})^*\,,
\end{equation}
such that the matrix-functions (\ref{Splus1}) and (\ref{Smin1}) become proportional to the identity matrix having the form
\begin{eqnarray}
i\Sigma^{(\pm)}(t,t',{\bf x})=\pm\frac{1}{(2\pi)^3}I_{\pm \epsilon}(t,t',{\bf x})\,,
\end{eqnarray}
which can be interpreted as the commutator functions of the Klein-Gordon field \cite{BDR}. Note that this happens only in the flat case since, in general, the matrix-functions $\Sigma^{\pm}$ are diagonal but are not proportional with the identity matrix. 
 
 { \begin{figure}
  \centering
    \includegraphics[scale=0.40]{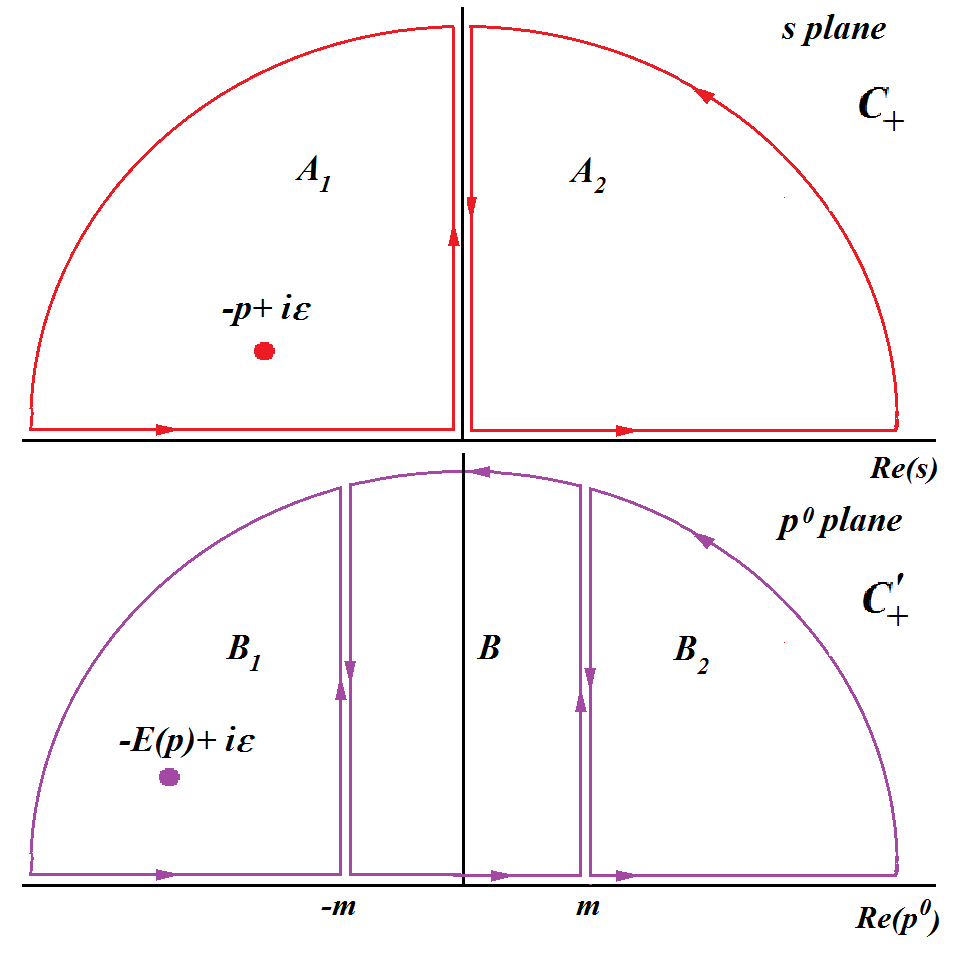}
    \caption{The contours of integration involved in changing  the integration variable: $C_{+}$, in the complex $s$-plane, and $C_+'$ in the complex plane $p^0$. }
  \end{figure}}

A problem could be here since  our integral rep. (\ref{SF1})  differs formally from the usual Fourier rep. of the Feynman propagator in Minkowski spacetime. In fact both these reps. are equivalent since these are related through a simple change of variable of integration, as we demonstrate in what follows. For proving this we start with the solutions of the system (\ref{sys}) which have the form 
\begin{equation}
w_s^{\pm}(t)=\sqrt{\frac{p^0(s)\pm m}{2p^0(s)}}\,e^{- i  p^0(s) t}
\end{equation}
where  now we denote $p^0(s)={\rm sign}(s) \sqrt{s^2+m^2}$ observing that
\begin{equation}\label{dom}
p^0(s)\in \left\{
\begin{array}{lll}
 [m,\infty)& {\rm if}& s\in [0,\infty)\,,\\
 (-\infty,-m]& {\rm if}&s\in(-\infty, 0]\,.
\end{array}\right.
\end{equation}
By substituting these functions in Eq. (\ref{SF1}) we obtain the integral rep.
\begin{eqnarray}
&&{S}_{F}(t,t^{\prime},{\bf x})
\nonumber\\
&&=\frac{1}{(2\pi)^4}\int {d^3p}e^{i{\bf p}\cdot{\bf x}}\int_{-\infty}^{\infty}ds\,\frac{s}{p^0(s)}\frac{\gamma^0p^0(s)-\gamma^i p^i +m}{s^2-p^2+i\epsilon}\,e^{-i p^0(s)(t-t')}\,,\nonumber\\\label{SFM} 
\end{eqnarray} 
where we may change the last integration variable $s\to p^0$. The problem is the integration domain since the codomain of the function $p^0(s)$ is $(-\infty,-m]\cup[m,\infty)$, as in Eq. (\ref{dom}), instead of ${\Bbb R}$. Considering the carresponding contour integrals, for example for $t'>t$, we may write the last integral as
\begin{equation}
{\cal I}=\int_{-\infty}^{\infty}ds...=\int_{C^+}ds...=\int_{A_1}ds...+\int_{A_2}ds...  
\end{equation}
where the contours $A_1$ and $A_2$ are shown in Fig. 2. After changing the variable we observe that 
\begin{equation}
\int_{A_{1,2}}ds..=\int_{B_{1,2}}dp^0\, \frac{p^0}{s}...\,,
\end{equation}
which means that we may write
\begin{eqnarray}
{\cal I}&=&\int_{B_{1}}dp^0\, \frac{p^0}{s}...+\int_{B_{2}}dp^0\, \frac{p^0}{s}...+\int_{B}dp^0\, \frac{p^0}{s}... \nonumber\\
&=&\int_{C'_{+}}dp^0\, \frac{p^0}{s}...=\int_{-\infty}^{\infty}dp^0\, \frac{p^0}{s}...
\end{eqnarray}
since the contour $B$ is sterile, giving $\int_{\small B}...=0$, as long as there are no poles in the domain $-m<p^0<m$. Thus we demonstrate that changing the variable $s\to p^0$ in our integral rep. (\ref{SFM}) we recover the well-known Fourier integral of the Feynman propagator of the massive Dirac fermions on Minkowski spacetime.   

\subsection{de Sitter expanding universe}

The de Sitter expanding universe is defined as the portion of the de Sitter manifold where the scale factor $a(t)=\exp(\omega t)$ depends  on the de Sitter Hubble constant denoted here by $\omega$ \cite{BD}. Consequently, in the conformal chart we have
\begin{equation}
t_c=-\frac{1}{\omega}e^{-\omega t}\in (-\infty, 0]~~~ \to~~~ a(t_c)=-\frac{1}{\omega t_c}\,.
\end{equation}
In these charts the normalized solutions of the system (\ref{sy1}) or (ref{sy2}) can be derived easily obtaining either their form in the FLRW chart 
\begin{equation}
u^{\pm}_p(t)=\sqrt{\frac{p}{\pi\omega}}\, e^{-\frac{1}{2}\omega t}K_{\nu_{\mp}}\left(-i\frac{p}{\omega}e^{-\omega t}\right)
\end{equation}
or the simpler expressions in the conformal chart
\begin{equation}
u^{\pm}_p(t_c)=\sqrt{-\frac{p t_c}{\pi}}\, K_{\nu_{\mp}}\left(i p t_c\right)
\end{equation}
where $K_{\nu_{\pm}}$ are the modified Bessel functions \cite{NIST} of the orders $\nu_{\pm}=\frac{1}{2}\pm i\frac{m}{\omega}$. The functions $v_p^{\pm}$ result from Eq. (\ref{VU}). Note that the normalization constant is derived from Eq. ({\ref{H3}).

Now we can write the  matrix-functions $\Sigma^{(\pm)}$ which depend on the basic integral  (\ref{Itt}) that now, after introducing the convergence parameter $\epsilon$ and using Eq. (6.692-2) of Ref. \cite{GR},  takes the form \cite{KP}
\begin{eqnarray}
&&I_{\epsilon}(t_c,t_c',{\bf x})=\frac{4\pi}{|{\bf x}|}\int_0^{\infty}dp\, p K_{\nu_{+}}(\epsilon p+ipt_c)K_{\nu_{+}}(-ipt_c') \sin p|{\bf x}|\nonumber\\
&&=\frac{\pi^2}{2(t_c t_c')^{\frac{3}{2}}}\,\Gamma\left(\textstyle{\frac{3}{2}}+\nu_{+}\right)\Gamma\left(\textstyle{\frac{3}{2}}-\nu_{+}\right) F\left[\textstyle{\frac{3}{2}}-\nu_{+},\frac{3}{2}+\nu_{+};2;1+\chi_{\epsilon}(t_c,t_c',{\bf x})\right]\,,\nonumber\\\label{Int1}
\end{eqnarray}
where the quantity  
\begin{equation}\label{chi}
\chi_{\epsilon}(t_c,t_c',{\bf x})=\frac{(t_c-t_c'-i\epsilon)^2-{\bf x}^2}{4t_ct_c'}\,,
\end{equation}
is related to the geodesic distance between the points $(t_c,{\bf x})$ and $(t_c', 0)$ \cite{BD}. Thus the structure of the matrix-functions $\Sigma^{(\pm)}$ and implicitly $S^{(\pm)}$ is completely determined. For example, the matrix-function (\ref{Splus1}) can be written in a closed form,
\begin{eqnarray}
&&i{\Sigma}^{(+)}_{\epsilon}(t_c,t_c^{\prime},{\bf x})=\frac{\omega^3}{16 \pi^2}\sqrt{t_ct_c'}\nonumber\\
&&\times\left\{
\pi_+ \Gamma\left(\textstyle{\frac{3}{2}}+\nu_{+}\right)\Gamma\left(\textstyle{\frac{3}{2}}-\nu_{+}\right)  F\left[\textstyle{\frac{3}{2}}-\nu_{+},\frac{3}{2}+\nu_{+};2;1+\chi_{\epsilon}(t_c,t_c',{\bf x})\right]\right.\nonumber\\
&&\left.+\pi_- \Gamma\left(\textstyle{\frac{3}{2}}+\nu_{-}\right)\Gamma\left(\textstyle{\frac{3}{2}}-\nu_{-}\right) F\left[\textstyle{\frac{3}{2}}-\nu_{-},\frac{3}{2}+\nu_{-};2;1+\chi_{\epsilon}(t_c,t_c',{\bf x})\right]\right\}\,,\nonumber\\\label{Splusf}
\end{eqnarray}
recovering thus the result of Ref. \cite{KP} for $D=4$ and $a=(-\omega t_c)^{-1}$. Moreover, the matrix-function ${\Sigma}^{(-)}_{\epsilon}$ can be derived from Eq. (\ref{SiSi}) as 
\begin{equation}
{\Sigma}^{(-)}_{\epsilon}(t_c,t_c^{\prime},{\bf x})=-{\Sigma}^{(+)}_{\epsilon}(t^{\prime}_c,t_c,{\bf x})
=-{\Sigma}^{(+)}_{-\epsilon}(t_c,t^{\prime}_c,{\bf x})\,,
\end{equation} 
since the expression (\ref{Splusf}) is symmetric in $t_c$ and $t_c'$ except $\chi_{\epsilon}$ for which the change $t_c\leftrightarrow t_c'$ reduces to $\epsilon \to -\epsilon$. Finally, the matrix-functions $S^{(\pm)}$ have to be calculated according to Eqs. (\ref{SSig}).

The integral rep. of the Feynman propagator of the massive Dirac field on this spacetime was derived in Ref. \cite{CIR} where we proposed for the first time the integral rep. studied here but with different notations. Now, starting with the solutions of the system (\ref{sys}) in the conformal chart,
\begin{equation}
w^{\pm}_s(t_c)=\sqrt{-\frac{s t_c}{\pi}}\, K_{\nu_{\mp}}\left(i s t_c\right)\,,
\end{equation} 
that have to be substituted in Eq. (\ref{SF1}), we recover the final result of Ref. \cite{CIR} in the present notations.

\subsection{A Milne-type spacetime}

The Milne universe \cite{BD} was intensively studied related to big-bang models but the Dirac field on this spacetime is less studied such that one knows so far only the exact solutions in a chart with spherical symmetry \cite{Sis}. Our approach allows us to study the Dirac field in a very close model of a spatially flat FLRW spacetime having a Milne-type scale factor $a(t)=\omega  t$ where $\omega$ is a free parameter. The principal difference is that this spacetime is no longer flat as the original Milne's universe, being produced by gravitational sources proportional with $\frac{1}{t^2}$.  

On this manifold, it is convenient to use the chart of proper time $\{t,{\bf x}\}$  for $t>0$. In this chart and the diagonal tetrad gauge (\ref{tetrad}), the system (\ref{sy1}) can be analytically solved finding  the solutions  
\begin{equation}\label{umiln}
u_p^{\pm}(t)=\sqrt{\frac{m}{2\pi}}\sqrt{t} \left[K_{\nu_+(p)}(i m t)\pm K_{\nu_-(p)}(imt)\right]\,,
\end{equation}
where the modified Bessel functions have the orders $\nu_{\pm}(p)=\frac{1}{2}\pm i \frac{p}{\omega}$.  These solutions comply with the normalization condition  (\ref{uuvv})  imposed with the help of the identity  (\ref{H3}) with $\mu={p}$. The corresponding functions 
\begin{equation}
v^{\pm}_p(t)= \sqrt{\frac{m}{2\pi}}\sqrt{t} \left[K_{\nu_-(p)}(-i m t)\mp K_{\nu_+(p)}(-imt)\right]\,,
\end{equation}
result from Eq. (\ref{VU}). Thus we obtain all the terms we need for building  the fundamental spinors (\ref{U}) and (\ref{V}) which could represent a new result even though it is somewhat elementary.  

The form of these functions is special since these are expressed in terms of modified Bessel functions which depend on $p$ through the orders instead of  arguments, as it happens in the de Sitter case. This leads to major difficulties in calculating the mode integrals since the basic integral (\ref{Itt}) cannot be solved by choosing a suitable formula from a table of integrals, requiring thus a special study.

However, despite of this impediment, we can write the integral rep. of the Feynman propagator by substituting the solutions of the system (\ref{sys}) that read,  
\begin{eqnarray}
w_s^{\pm}(t)&=&\sqrt{\frac{m}{2\pi}}\sqrt{t} \left\{K_{\nu_+(s)}[{\rm sign}(s) i m t]\right.\nonumber\\
&&\left.\pm {\rm sign}(s)K_{\nu_-(s)}[{\rm sign}(s)imt]\right\}\,,\label{umiln}
\end{eqnarray}
in Eq. (\ref{SF1}). Thus we obtain the principal piece we need for calculating Feynman diagrams in this spacetime.

\section{Concluding remarks}

We presented here the complete theory of the free Dirac fermions on spatially flat FLRW spacetimes including the expressions of the propagators in the configuration rep. as integrals encapsulating the effect of the Heaviside step functions. In this approach we have only two scalar modulation functions $u^{\pm}_p(t)$ which depend on the concrete geometry. These have to be derived solving the systems (\ref{sy1}) or (\ref{sy1c}) or resorting to numerical integration on computer. In this manner any Feynman diagram involving fermions can be calculated either analytically or numerically. 

We have thus all the elements we need for developing the QED on spatially flat FLRW spacetime since the theory of the Maxwell field in the conformal charts is similar to that of special relativity since the Maxwell equations are conformally  invariant. The only difficulty is the gauge fixing which, in general,  does not have this property except the Coulomb gauge which becomes thus mandatory  \cite{CQED}.  

Concluding we can say that the approach presented here offers us the integral representation of the fermion propagators which are the missing pieces we need for building a coherent QED which could be the core of a future QFT on FLRW spacetimes. 

\appendix

\section{Pauli and Dirac spinors}

The Pauli spinors depend on the direction of the spin projection \cite{BDR,TH}. Thus in the  momentum-spin basis  where the spin is projected on the third axis of the rest frame  the particle spinors  $\xi_{\sigma}$ of polarization $\sigma=\pm\frac{1}{2}$ are $\xi_{\frac{1}{2}}=(1,0)^T$ and $\xi_{-\frac{1}{2}}=(0,1)^T$ while the antiparticle ones  are defined as $\eta_{\sigma}=i\sigma_2 \xi_{\sigma}$  \cite{TH}. These spinors  satisfy the eigenvalues problems $S_3 \xi_{\sigma}=\sigma \xi_{\sigma}$ and $S_3 \eta_{\sigma}=-\sigma \eta_{\sigma}$, where $S_i=\frac{1}{2}\sigma_i$ are the spin operators in terms of Pauli's matrices, and  are  normalized correctly,  $\xi^+_{\sigma}\xi_{\sigma'}=\eta^+_{\sigma}\eta_{\sigma'}=\delta_{\sigma\sigma'}$,  satisfying the completeness condition 
\begin{equation}\label{Pcom}
\sum_{\sigma}\xi_{\sigma}\xi_{\sigma}^+=\sum_{\sigma}\eta_{\sigma}\eta_{\sigma}^+={\bf 1}_{2\times 2}\,.
\end{equation}

The corresponding normalized Pauli spinors of the momentum-helicity basis, $\xi_{\lambda}({\bf p})$, and  $\eta_{\lambda}({\bf p})$,  of helicity $\lambda =\pm\frac{1}{2}$,   satisfy the eigenvalues problems $({\bf p}\cdot {\bf S})\,\xi_{\lambda}({\bf p})=\lambda\, p\, \xi_{\lambda}({\bf p})$ and respectively $({\bf p}\cdot {\bf S})\,\eta_{\lambda}({\bf p})=-\lambda\, p\, \eta_{\lambda}({\bf p})$, having the form \cite{BDR}
\begin{eqnarray}
\xi_{\frac{1}{2}}({\bf p})&=&\sqrt{\frac{p+p^3}{2p}}\left(
\begin{array}{c}
1\\
\frac{p^1+i p^2}{p+p^3}
\end{array}\right)\,,\\ 
\xi_{-\frac{1}{2}}({\bf p})&=&\sqrt{\frac{p+p^3}{2p}}\left(
\begin{array}{c}
\frac{-p^1+i p^2}{p+p^3}\\
1
\end{array}\right)\,,
\end{eqnarray}
and $\eta_{\lambda}({\bf p})=i\sigma_2 \xi_{\lambda}({\bf p})^*$.
These spinors comply with  completeness conditions similar to Eqs. (\ref{Pcom}) such that we may construct the projectors  \cite{TH}
\begin{eqnarray}
P_{\lambda}&=&\xi_{\lambda}({\bf p})\xi_{\lambda}({\bf p})^+=\eta_{-\lambda}({\bf p})\eta_{-\lambda}({\bf p})^+\nonumber\\
&=&\frac{1}{2}\,{\bf 1}_{2\times 2}+\lambda\,\frac{{\bf \sigma}\cdot{\bf p}}{p}\,,
\end{eqnarray}
that can be used when the helicities of the massless fermions are restricted as in the case of the left-handed neutrinos whose system of spinors is incomplete. 

Here we use the Dirac matrices $\gamma^{\hat\alpha}$ which are self-adjoint, $\bar{\gamma}^{\hat\alpha}=\gamma^0 (\gamma^{\hat\alpha})^+\gamma^0=\gamma^{\hat\alpha}$, and the matrix $\gamma^5=i\gamma^0\gamma^1\gamma^2\gamma^3$ which is anti self-adjoint, $\bar{\gamma}^5=-\gamma^5$. For these matrices we use either the standard rep.  (with diagonal $\gamma^0$) or the chiral rep.  where $\gamma^5$ and the left and right-handed projectors, 
\begin{equation}\label{PLR}
P_L=\frac{1-\gamma^5}{2}\,, \qquad P_R=\frac{1+\gamma^5}{2}\,,
\end{equation}
are diagonal \cite{TH}. 

The charge conjugation $\psi\to\psi^{c}=C\bar\psi^{T}$  is defined by the matrix $C=i\gamma^{2}\gamma^{0}$ which  commutes with $\gamma^5$ and satisfies $\gamma^{\hat\mu}C=-C(\gamma^{\hat\mu})^T$. Hereby, we may deduce simple calculation rules as $(\gamma^{\hat\alpha})^c=C(\gamma^{\hat\alpha})^TC^{-1}=-\gamma^{\hat\alpha}$  and  verify that the charge conjugation changes the  chirality,
\begin{equation}
(\psi_{L/R})^{c}=\frac{1\pm\gamma^{5}}{2}\,\psi^{c}=(\psi^{c})_{R/L}\,,
\end{equation}
since $(\gamma^5)^c=-\gamma^5$ \cite{BDR,TH}.

The rest frame spinors of the momentum-spin basis are given by Eqs. (\ref{Rfspin}) where the Pauli spinors of the spin basis can be replaced at any time by those of the helicity basis if we want to work in the momentum-helicity basis (where we do not have rest frames). In contrast, the rest frame spinors in the chiral rep. have the form \cite{BDR,TH}
\begin{equation}
u_{\sigma}=\frac{1}{\sqrt{2}}\left(
\begin{array}{l}
\xi_{\sigma}\\
\xi_{\sigma}
\end{array}\right) \quad
v_{\sigma}=-C\bar{u}_{\sigma}^T=\frac{1}{\sqrt{2}}\left(
\begin{array}{c}
-\eta_{\sigma}\\
\eta_{\sigma}
\end{array}\right)
\end{equation}
and similarly for those of the momentum-helicity basis. 

\section{The modified Bessel functions $K_{\nu_{\pm}}(z)$}

According to the general properties of the modified Bessel functions, $I_{\nu}(z)$ and $K_{\nu}(z)=K_{-\nu}(z)$ \cite{NIST},  with
$\nu_{\pm}=\frac{1}{2}\pm i \mu$  ($ \mu \in {\Bbb R}$), are related among themselves through
\begin{equation}\label{H1}
[K_{\nu_{\pm}}(z)]^{*}
=K_{\nu_{\mp}}(z^*)\,,\quad \forall z \in{\Bbb C}\,,
\end{equation}
satisfy the equations
\begin{equation}\label{H2}
\left(\frac{d}{dz}+\frac{\nu_{\pm}}{z}\right)K_{\nu_{\pm}}(z)=-K_{\nu_{\mp}}(z)\,,
\end{equation}
and the identities
\begin{equation}\label{H3}
K_{\nu_{\pm}}(z)K_{\nu_{\mp}}(-z)+ K_{\nu_{\pm}}(-z)K_{\nu_{\mp}}(z)=\frac{i\pi}{ z}\,,
\end{equation}
that guarantees the correct orthonormalization properties of the fundamental spinors. For 
$|z|\to \infty$  we have  \cite{NIST}
\begin{equation}\label{Km0}
I_{\nu}(z) \to \sqrt{\frac{\pi}{2z}}e^{z}\,, \quad K_{\nu}(z) \to K_{\frac{1}{2}}(z)=\sqrt{\frac{\pi}{2z}}e^{-z}\,,
\end{equation} 
for any $\nu$.

\subsection*{Acknowledgments}

This work is partially supported by a grant of  the Romanian Ministry of Research and Innovation, CCCDI-UEFISCDI, project number  PN-III-P1-1.2-PCCDI-2017-0371.

\end{document}